\documentstyle[12pt,epsf]{article}
\begin{document}

\begin{center}
\begin{Large}
{\bf Binary Evolution in World Wide Web}

\vskip 2cm

\end{Large}
S.N.Nazin$^1$, V.M.Lipunov$^{1,2}$, I.E.Panchenko$^2$,
K.A.Postnov$^2$, M.E.Prokhorov$^1$ and S.B.Popov$^2$

\vskip 1cm

$^1$ Sternberg Astronomical Institute

$^2$ Department of Physics, Moscow State University

\vskip 2cm

{\bf Abstract}

\end{center}
\vskip 0.5cm

We present a WWW-version of the {\it Scenario Machine} - a computer
code designed to calculate the  evolution of
close binary stellar systems. The Internet users
can directly access to the code and calculate binary evolutionary
tracks with parameters at the user's will.  The program is running on
the {\it Pentium} server of the Division of the Relativistic
Astrophysics of the Sternberg Astronimical Institute ({\tt
http://xray.sai.msu.su/ }). The results are presented both in the form of
tables and graphic diagrams.  The work is always in progress.  More
possibilities for Internet users are intended to become available in
the near future.

\clearpage

\section{Introduction}

The Scenario Machine method to calculate evolution of
binary stars is basically
a Monte-Carlo method for statistical simulation of large
ensembles of stellar binary systems originally used by
Kornilov \& Lipunov (1983) for massive binaries
and developed later by Lipunov \& Postnov (1987) for low-mass binaries
(for the most recent review, see Lipunov, Postnov \&Prokhorov, 1996 (LPP96)).
The method is based on the construction
of a great number of single evolutionary tracks
with different initial conditions.

Although the present WWW-version of the Scenario Machine includes
only the single-track constructor,
it allows one to make quickly
the evolutionary track of a close binary system with arbitrary
initial parameters, such as nmasses of the components, orbital separation,
orbital eccentricity, magnetic fields of compact stars, etc.,
as well as using different types of evolutionary scenarios
(with different distributions of the kick velocity
imparted to a newborn neutron star, high- or low-mass stellar wind loss
diring main sequence evolution, etc.).

This is not the first example of the WWW
representation of the stellar evolution.
In 1995,
the Goettingen group made available direct calculations of single stars
evolution  with parameters determined by the user on their
WWW-server ({\tt http://www.uni-sw.gwdg.de/~jloxen/wwwgal/}).
In contrast, our code affords for the first time the calculation of {\it binary}
evolutionary tracks.

\section{The Model}

In the calculations of the evolutionary track both
nuclear evolution of the normal star and spin evolution of a
magnetized compact object (neutron star (NS) or white dwarf (WD)) are
taken into account.

\subsection{Normal star evolution}

We consider stars with a constant (solar) initial chemical composition.
Before the Roche lobe overflow, the components evolve as single
stars (see for details LPP96).
Analytical approximation describing evolutionary tracks are used.
For massive stars, we use two types of evolutionary tracks:
those calculated by the Geneva group
(Schaller et al., 1992) with high stellar wind mass-loss
(up to 90 per cent of the initial mass), and
those calculated earlier by different groups
assuming low stellar wind mass-loss (up to 30 per cent of the initial mass).

When a star fills its Roche lobe,
the mass transfer time-scales are treated differently depending on
the star's mass and time it fills the Roche lobe.
For the most close binaries we aslo calculate
Roche lobe overfilling due to angular momentum loss by
magnetic stellar wind or gravitational radiation.

If the mass transfer onto a normal star occurs on a time-scale ten time
shorter than the thermal Kelvin-Helmholz time for this star or a
compact star is engulfed by a giant companion (e.g. as a result of the
kick), the common envelope (CE) stage of the binary evolution is set in
(Paczynski 1976). The CE stage is treated conventionally by by
introducing a parameter $\alpha_{CE}$ that measures what fraction of
the system's orbital energy goes, between the beginning and the end of
the spiralling-in process, into the binding energy (gravitational minus
thermal) of the ejected common envelope.  Spiral-in during the CE stage
can result in the binary coalescence.  In the case of coalescence of
the normal star with a compact object (NS or BH) a Thorne-Zytkow object
is formed.

\subsection{Initial parameters of compact stars}

The stars with initial masses $M\le 10 M_\odot$
leave a WD in the end of evoluiton, with the WD masses
and chemical composition depending on the binary system's
parameters. Stars with $10 M_\odot<M<M_{cr}$ collapses
to form NS of 1.4 M$_\odot$; when $M>M_{cr}$ a BH with
a mass of $M_{bh}=k_{bh}M_*$ is formed, where
$M_*$ is either the mass of the pre-collapsing star
(in the case of low stellar wind mass-loss scenario), or
the initial mass of the star (in teh case of the high stellar wind
mass-loss scenario). NS and WD are assumed to have
a magnetic field randomly distributed within
a range $10^{5}-10^{9}$ G and $10^{8}-10^{13}$ G
correspondingly.

The evolution of a compact star
is considered as the change of its spin period and
hence the change of regime of interaction with the surrounding
plasma supplied by the second component (for more detail
see Lipunov 1992, Lipunov \& Popov 1995).

The accretion rate is limited by the Eddington luminosity or
by the surface nuclear burning of the accreted matter (for WD).
In that case supercritical regimes with mass outflow may happen.

When an accreting WD reaches
the Chandrasekhar limit, it assumes to explode as a
SN type Ia with or without NS formation.  Analogously, after reaching
the Oppengeimer-Volkov limit, a NS collapses into a BH.

\section{World Wide Web version}

The World Wide Web version provides the opportunity to construct an
evolutionary track with all important parameters determined
by the user at will from a broad range of available values.


Some of the most unclear branches of the scenario can also be
determined by the user.
They include: the type of mass-loss of normal star,
the lewel of conservativeness of the matter captured
by the compact companion,
the upper limit of the accretion rate at the CE stage by the compact object,
switching of the accretion-induced magnetic field decay, etc.

Below we describe briefly how to handle the code,
what the WWW-pages contain and how the results are presented.


\subsection{WWW access}

 The code is located on the WWW-server of the Division of Relativistic
Astrophysics of the Sternberg Astronomical Institute. The URL is \\
 {\tt http://xray.sai.msu.su/sciwork/scenario.html}.
The server quickly calculate the track so the waiting time is
short.

\subsection{Page contents}

On the main page, a very short description of the program may be found.

From the main page three following items are available:

1. "Go to the evolutionary track constructor"

2. "How all this stuff work?"

3. "Credits"

The first link brings the user to a very friendly interface of the main
code, where all parameters can be set and the calculations can be started.

The second lonk provides a
list of papers containing the results of the Scenario Machine
calculations beginning
from 1983 untill 1996 and a short description of the code.

The third link  gives some information on the authors of the code
and its WWW-version.

\subsection{Presentation of the results}

The results are presented in  forms of tables and diagrams chosen by
the user at will.  Part of the table and a cartoon figure of the track
calculated for the default set of the initial parameters are shown in
Fig. 1 and 2.

In the future, we intend to made available calculations
of a large number of binary systems assuming different
intial distributions, chemical composition, etc., to
construct a "Galaxy at will", as well as various
interface facilities.

\section{Acknowledgements}

 We acknowledge the
 Geneva group for comments on their evolutionary tracks.

The work was partially supported by the INTAS grant No 93-3364, grant
of Russian Fund for Basic Research No 95-02-06053a, grant JAP-100 from
the International Science Foundation and Russian Government and by
Center for Cosmoparticle Physics ``COSMION'' (Moscow, Russia).

\end{document}